\documentclass[12pt]{article}
\usepackage{latexsym}
\usepackage{epsfig}
\def\bseq{\begin{subequation}}  % = 1a 1b
\def\eseq{\end{subequation}}
\def\bsea{\begin{subeqnarray}}  % = 1.1a 1.1b
\def\esea{\end{subeqnarray}}
\def\bc{\begin{center}}
\def\ec{\end{center}}

\newcommand{\beq}{\begin{equation}}
\newcommand{\eeq}{\end{equation}}
\newcommand{\bea}{\begin{eqnarray}}
\newcommand{\eea}{\end{eqnarray}}
\newcommand{\bdm}{\begin{displaymath}}
\newcommand{\edm}{\end{displaymath}}
\newcommand{\ba}{\begin{array}}
\newcommand{\ea}{\end{array}}
\newcommand{\ben}{\begin{enumerate}}
\newcommand{\een}{\end{enumerate}}
\newcommand{\bde}{\begin{description}}
\newcommand{\ede}{\end{description}}
\newcommand{\nn}{\nonumber}

\renewcommand{\r}{\right}
\renewcommand{\l}{\left}

\newcommand{\qq}{\qquad}

\def\bc{\begin{center}}
\def\ec{\end{center}}

\def\bfg{\begin{figure}}
\def\efg{\end{figure}}
\newcommand{\Ci}{{\ \hbox{{\rm I}\kern-.6em\hbox{\bf C}}}}
\newcommand{\erre}{{\hbox{{\rm I}\kern-.2em\hbox{\rm R}}}}
\newcommand{\1}{ \,  \raisebox{+0.14em}{{\hbox{{\rm \scriptsize ]}}
\raisebox{-0.2em}{\kern-.8em\hbox{1}}}} \, }
\newcommand{\oc}{{\big <}}
\newcommand{\cc}{{\big >}}
\def\Hat#1{\widehat{#1}}  % big hat
\newcommand{\bbox}{\lower.2ex\hbox{$\Box$}}

\newcommand{\Tr}{{\rm Tr}}        % traccia
\newcommand{\STr}{{\rm STr}}      %supertraccia
\newcommand{\pa}{\partial}        %derivata parziale
\newcommand{\ub}[1]{\underbrace{#1}}

\renewcommand{\a}{\alpha}

\renewcommand{\b}{\beta}

\newcommand{\C}{\Gamma}
\renewcommand{\d}{\delta}

\newcommand{\ep}{\epsilon}

\renewcommand{\th}{\theta}

\newcommand{\la}{\lambda}

\newcommand{\m}{\mu}

\newcommand{\n}{\nu}

\newcommand{\f}{\phi}

\newcommand{\s}{\sigma}
\renewcommand{\S}{\Sigma}

\newcommand{\DC}{\nabla}
\newcommand{\DCb}{\overline{\nabla}}

\newcommand{\ad}{\dot{\alpha}}
\newcommand{\bd}{\dot{\beta}}

\newcommand{\Db}{\olD}
\newcommand{\Wb}{\olW}

\newcommand{\olW}{\overline{W}}
\newcommand{\olcW}{\overline{\cal W}}
\newcommand{\fb}{\bar{\phi}}
\newcommand{\olD}{\overline{D}}

\newcommand{\olth}{\overline{\theta}}

\newcommand{\st}{\tilde{\sigma}}

\newcommand{\cW}{{\cal W}}

\newcommand{\cN}{{\cal N}}

\begin{document}
\begin{titlepage}
\begin{flushright}
IFUM-659/FT \\
KUL-TF-2000/18\\
%hep-th/0006067
\end{flushright}
\vspace{.3cm}
\begin{center}
{\Large \bf Non abelian ${\cal N}=2 $ supersymmetric\\
Born Infeld action}
\vfill%\vskip 15mm%27.mm

{\large \bf Andrea Refolli$^1$, Niccolo' Terzi$^2$ and
Daniela Zanon$^2$}\\
\vfill%\vskip 7mm%1cm

{\small
$^1$  Instituut voor Theoretische Fysica - Katholieke Universiteit Leuven
\\Celestijnenlaan 200D B--3001 Leuven, Belgium\\
\vspace*{0.4cm}
$2$ Dipartimento di Fisica dell'Universit\`a di Milano
and\\ INFN, Sezione di Milano, Via Celoria 16,
20133 Milano, Italy\\}
\end{center}
\vfill
\begin{center}
{\bf Abstract}
\end{center}
{\small We present a ${\cal N}=2$ supersymmetric action for the Born
Infeld theory in the non abelian case. We quantize the theory in
${\cal N}=1$ superspace and compute divergences at one-loop.
The result is discussed in the ${\cal N}=4$ case.}

\vspace{2mm} \vfill \hrule width 3.cm
\begin{flushleft}
e-mail: andrea.refolli@fys.kuleuven.ac.be\\
e-mail: terzi@pcteor1.mi.infn.it\\
e-mail: daniela.zanon@mi.infn.it
\end{flushleft}
\end{titlepage}

 \section{Introduction}

The Born Infeld theory describes the open string tree level
 effective action in
the approximation of slowly varying field strengths \cite{GW,T1}.
It also appears in the analysis of the low-energy dynamics of a
D-brane \cite{L}. In both cases it is sufficient to consider an abelian
theory in terms of a $U(1)$ gauge field. When many D-branes are put
together and let coincide the gauge group becomes $U(N)$ \cite{W1}.
The Born Infeld action to be studied becomes correspondingly a non abelian
one. The replacement of the abelian field strength $F_{\m\n}$ by a
non abelian tensor field is not uniquely defined.  The problem
 has been extensively studied and discussed \cite{many,T2,HT,BP,T3,denef}. 
Essentially it has to
do with the ordering ambiguity in the group trace operation and in
addition with the fact that derivatives acting on the field strength
cannot be completely separated out since 
$[D_{\rho},D_\s] F_{\m\n}=[F_{\rho\s},F_{\m\n}]$.

The proposal in ref. \cite{T2} is the simplest non abelian extension
of the abelian Born Infeld action. For its intrinsic elegance 
it is appealing by itself; in fact it exhibits
 also many other advantages.
In particular it exactly matches the full non abelian open
string effective action
at least up to order $\a'^2$, i.e. it correctly reproduces
the terms $F^2+\a'^2F^4$. It might represent a good approximation
for a string dynamics where nearly covariantly constant
field strengths are relevant.

The abelian bosonic Born Infeld theory admits supersymmetric ${\cal N}=1$
 and ${\cal N}=2$
versions.
 A non abelian  ${\cal N}=1$ supersymmetrization of the theory has
been proposed in ref. \cite{GSS}: it has been defined in such a way
that the bosonic part of the action reduces to the one in \cite{T2}.

In this paper we extend the construction presented in \cite{GSS,GNSS} to
the ${\cal N}=2$ non abelian Born Infeld theory. We do that in terms
of ${\cal N}=2$ superfields which allow an almost direct and
straightforward generalization of the ${\cal N}=1$ case. Indeed using
the results in ref. \cite{ketov1}, and rewriting appropriately the
formulas obtained in \cite{GSS} we reach our goal most easily.
Then we consider the quartic interaction term, express it in
terms of ${\cal N}=1$ superfields and check its bosonic content. We
show that it matches the $F^4$ terms in the bosonic non
abelian action of ref. \cite{T2}.

We proceed in analogy with what has been done in \cite{DSZ} for the abelian
${\cal N}=2$ theory: we perform the quantization in ${\cal N}=1$
superspace and consider ${\cal O}(\a'^4)$ one-loop
corrections to the on-shell effective action. We determine the
structure of the counterterm which is proportional to derivatives of
the field strength. Finally, even if the complete Born Infeld action
is not known for the ${\cal N}=4$ case, nonetheless the
 ${\cal O}(\a'^4)$ one-loop result is easily computable. It can
 be written in a very symmetric and elegant form. Since the
 corresponding contribution in the abelian theory was
 consistent with effective action calculations from superstring
 theory, it is suggestive to expect the non abelian result
 be in accordance with super $D$-brane dynamics.

\vspace{0.8cm}

We work in superspace following the notations and conventions in ref.
\cite{superspace} and \cite{WGR}.  First we briefly review the construction
of the Born Infeld action in the abelian case.\\
The ${\cal N}=1$ superfields of
interest are the chiral field strengths $W^\a$, $\bar{W}^{\ad}$, in
terms of which the supersymmetric action can be written as \cite{CF}
\bea
\label{N1abBI}
S_{BI}^{(1)} & = & \int {\rm d}^4x~{\rm d}\th^2 W^2 +
 \int{\rm d}^4x ~{\rm d}\bar{\th}^2 \Wb^2 +
 \int {\rm d}^4x~{\rm d}\th^2 {\rm d}\bar{\th}^2 W^\a W_\a
 \Wb^{\ad}\Wb_{\ad}~ {\cal B}(K,\bar{K}) \nn \\
&&~~~~
\eea
where
\beq
{\cal B}(K,\bar{K}) = \l[1 - \frac{K+\bar{K}}{2} +
\sqrt{1-(K+\bar{K})+\frac{1}{4}(K-\bar{K})^2}\r]^{-1}
\label{def1}
\eeq
and
\beq
K = D^2(W^\a W_\a)\qquad\qquad\qquad\bar K = \Db^2(\Wb^{\ad} \Wb_{\ad})
\label{def2}
\eeq
It is an easy matter to check that the bosonic part of this action just
reproduces the standard Born Infeld action. Indeed one can proceed as
follows: first one introduces
the field components of $W^\a$ defined as
\beq
\la_\a = W^{\a}\vert \qquad\qquad f_{\a\b}=
\frac{1}{2} D_{\l(\a\r.}W_{\l.\b\r)}\vert \qquad\qquad
D'=-\frac{i}{2} D^{\a}W_{\a} \vert
\label{N1components}
\eeq
where $\vert$ indicates setting $\theta^{\a}=\bar\theta^{\ad}=0$.
In particular using the definitions
\beq
(\s_\m)_{\a\ad} = (\1,\vec{\s}) \qquad\qquad\quad
(\st_\m)^{\ad\a} =(\1,-\vec{\s})
\eeq
and
\beq
(\s_{\m\n})_{\a}^{~\b}  \equiv
-\frac{1}{4}(\s_\m\st_\n-\s_\n\st_\m)_{\a}^{~\b} \qquad\qquad
(\st_{\m\n})^{\ad}_{~\bd}  \equiv
-\frac{1}{4}(\st_\m\s_\n-\st_\n\s_\m)^{\ad}_{~\bd}
\label{sigmamatrix}
\eeq
the electromagnetic antisymmetric tensor can be expressed in
terms of the two symmetric bispinors
\beq
\label{emt}
F_{\m\n} = (\s_{\m\n})_{\a\b} f^{\a\b} -(\st_{\m\n})_{\ad\bd} \bar f^{\ad\bd}
\eeq
Then with the definitions
\beq
F^2\equiv F^{\m\n}F_{\m\n}\qquad\qquad\qquad
 F^4 \equiv F_{\m\n}F^{\n\rho}F_{\rho\s}F^{\s\m}
\label{F2F4}
\eeq
one obtains
\bea
&~&-(K+\bar{K})\vert = \frac{1}{2}F^2\equiv X \nn \\
&~&(K - \bar{K})^2\vert =-F^4+\frac{1}{2}(F^2)^2\equiv -4Y^2\nn \\
&~& W^2\vert_F + \Wb^2 \vert_F =-\frac{1}{4}F^2\equiv I_2\nn\\
&~& \frac{1}{2}(W^\a W_\a \Wb^{\ad}\Wb_{\ad})\vert_D
=\frac{1}{8}(F^4 -\frac{1}{4}(F^2)^2)\equiv I_4
\label{N1invariants}
\eea
Finally in terms of the above invariants the bosonic Born Infeld
 lagrangian takes the form
\bea
L_{BI} & =&  \l(1-\sqrt{-det_4(\eta_{\m\n} + F_{\m\n})}\r)\nn\\
&=&I_2 + 2I_4 \l[ 1+\frac{X}{2} +\sqrt{1+X-Y^2}\r]^{-1}
\label{bosBIaction}
\eea
The expression in (\ref{bosBIaction}), with the identifications in
(\ref{N1invariants}), clearly coincides with the bosonic part of the
${\cal N}=1$ supersymmetric action in (\ref{N1abBI}).

\vspace{0.8cm}

The same approach allows one to obtain the ${\cal N}=2$ supersymmetric
abelian version of the action \cite{ketov1}\footnote{ Note however that beyond quartic interactions the form of the action is not unique, see for example
\cite{theisen}}.
The ${\cal N}=2$ superfields of interest are the chiral field
strengths
$\cW$ and $\olcW$ which satisfy the constraints
\beq
D^{\a}_a D_{b\a} \cW = C_{ac}C_{bd}\Db^{d\ad}\Db_{\ad}^c \olcW
\label{constraint}
\eeq
where $C_{ab}$ is the Levi-Civita antisymmetric tensor.
The $\cN =1$ components are defined as
\beq
\f \equiv \cW\vert\qquad\qquad\qquad
 W_\a \equiv - D_{2\a} \cW\vert
 \label{N2components}
\eeq
and everything is evaluated at $\theta_2^\a=\bar\theta^{\ad}_2=0$.
In complete correspondence with the equations in
(\ref{N1abBI}), (\ref{def1}) and (\ref{def2}) one has
\bea
S_{BI}^{(2)} & = & \frac{1}{2}\int {\rm d}^4x{\rm d}\th^4 \cW^2 +
\frac{1}{2} \int{\rm d}^4x {\rm d}\bar{\th}^4 \olcW^2 +
 \int {\rm d}^4x{\rm d}\th^4 {\rm d}\bar{\th}^4 ~\cW\cW\olcW\olcW~
 {\cal Y}(K,\bar{K}) \nn \\
&&~~~~
\label{N2abBI}
\eea
where
\beq
{\cal Y}({\cal K},\bar{\cal K}) = \l[1 - \frac{{\cal K}+\bar{\cal K}}{2}
+ \sqrt{1-({\cal K}+\bar{\cal K})+\frac{1}{4}
({\cal K}-\bar{\cal K})^2}\r]^{-1}
\label{def1N2}
\eeq
and
\beq
{\cal K}= D^4 \cW^2 \qquad\qquad\quad
\bar{\cal K}= \Db^4 \olcW^2
\label{def2N2}
\eeq

\vspace{0.8cm}

Now we turn to the construction for the non abelian case. The various
actions which appear in (\ref{bosBIaction}), (\ref{N1abBI}),
(\ref{N2abBI})
contain all order interaction terms obtained  by the power series expansion
of the square root. In order to promote these theories from abelian
to non abelian ones it is sufficient to introduce
gauge covariant derivatives,
treat $F^{\m\n}$, $W^\a$, ${\cal W}$
as matrices, expand the square root as before and take the
trace of the various terms. In order to overcome the ordering
ambiguity, in \cite{T2} it has been suggested to introduce a
symmetrized trace defined for any set of matrices $A_1,A_2,\dots,A_n$
as
\bea
\STr(A_1,A_2,\ldots,A_n) &=&
 \frac{1}{n!}\sum_{perm.} \Tr(A_{\s_1},A_{\s_2},\ldots,A_{\s_n})
\eea
For the bosonic action in (\ref{bosBIaction}) we then obtain the
non abelian generalization
\bea
L_{BI} & = & \STr \l(1-\sqrt{-det_4(\eta_{\m\n} + F_{\m\n})}\r) \nn \\
&=& \sum_{n=0}^{\infty} q_n \STr[(X-Y^2)^{n+1}]
=\sum_{n=1}^{\infty} q_{n-1} \sum_{j=0}^n {n \choose j} \STr[X^j (-Y^2)^{n-j}]
\label{bosnonabBI}
\eea
with
\beq
q_0 = -\frac{1}{2} \qquad\qquad\quad
q_n  =  \frac{(-1)^{n+1}}{4^n} \frac{(2n-1)!}{(n+1)! (n-1)!}
\label{qcoeff}
\eeq
Correspondingly in the ${\cal N}=1$ case, from (\ref{N1abBI}) we
write
\bea
\label{N1nonabBI}
S_{BI}^{(1)} & = & \int {\rm d}^4x~{\rm d}\th^2 ~\Tr W^2
 + \int{\rm d}^4x ~{\rm d}\bar{\th}^2~ \Tr\Wb^2   \\
&& + \sum_{n,m} \frac{C_{n,m}}{2}
\int {\rm d}^4x~{\rm d}\th^2 {\rm d}~\bar{\th}^2
~ \STr(W^\a, W_\a, \Wb^{\ad},\Wb_{\ad},\Hat{X}^n,\Hat{Y}^m) \nn
\eea
where
\bea
&&\Hat{X}= -(K+\bar K)\qquad\qquad\qquad ~~\Hat{X} \vert=X \nn\\
&&\Hat{Y} = -\frac{i}{2}(K - \bar K) \qquad\qquad\qquad
 \Hat{Y} \vert=Y
 \label{XYN1}
 \eea
The lowest order interaction is given by
\beq
L_4=\frac{1}{2}\STr(W^\a, W_\a, \Wb^{\ad},\Wb_{\ad})
= \frac{1}{3}\Tr (W^\a W_\a\Wb^{\ad}\Wb_{\ad} -
\frac{1}{2}W^\a\Wb^{\ad}W_\a\Wb_{\ad})
\label{4WWbar}
\eeq
whose bosonic expression exactly reproduces the non abelian bosonic
terms proposed in \cite{T2}. Indeed one obtains
\bea
L_{4~bos} &=& \frac{1}{3}\Tr (f^{\a\b}f_{\a\b}\bar{f}^{\ad\bd}
\bar{f}_{\ad\bd} - \frac{1}{2}f^{\a\b}\bar{f}^{\ad\bd}f_{\a\b}
\bar{f}_{\ad\bd}) \nn \\
&=& \frac{1}{12}\Tr(F^{\m\n}F_{\n\s}F_{\m\rho}F^{\rho\s}
 + \frac{1}{2}F^{\m\n}F_{\n\s}F^{\rho\s}F_{\m\rho} \nn \\
&~& \qq -\frac{1}{4} F^{\m\n}F_{\m\n}F^{\rho\s}F_{\rho\s}
-\frac{1}{8}F^{\m\n}F^{\rho\s}F_{\m\n}F_{\rho\s}) \nn \\
&=& \frac{1}{8} \STr( F^4 - \frac{1}{4}(F^2)^2)
\label{4bos}
\eea
Now one has to determine
the higher order terms which appear in (\ref{N1nonabBI}).
First of all one has to
consider the symmetrized trace of powers
of $n$, $m$ matrices $\Hat{X}^n$,  $\Hat{Y}^m$. Moreover for the
superfields $K$ e $\bar{K}$ contained in  $\Hat{X}$ e
$\Hat{Y}$ one has to use the decomposition
\bea
\DC^2 (W^\a W_\a) &=& (\DC^2 W^\a) W_\a + W^\a (\DC^2 W_\a)
-\DC^\a W^{\b} \DC_\a W_{\b}
\eea
and remember that the $\STr$ operation permutes each  $W_\a$ factor.

Then the coefficients $C_{n,m}$ need to be computed. Even if in
ref. \cite{GSS}, \cite{GNSS} they have been evaluated for the
${\cal N}=1$ theory, since we are using a different notation we give
here our derivation that immediately extends to the ${\cal N}=2$
case.
The problem is to compare the following two series
\bea
L_{BI} & = & \sum_{n=1}^{\infty} q_{n-1} \sum_{j=0}^n {n \choose j}
 X^{n-j} (-Y^2)^{j}\nn \\
&= & \sum_{n,m} (q_1 X^2 - q_0 Y^2) C_{n,m}X^n Y^m
 \label{eq:series}
\eea
We immediately have $C_{n,2m+1}=0$. Then since
 \mbox{$q_0=-\frac{1}{2},~q_1=\frac{1}{8}$} one can rewrite
\bea
L_{BI} & = &\sum_{n,m} (q_1 C_{n,2m} X^{n+2}(Y^2)^{m}
- q_0 C_{n,2m}X^n (Y^2)^{m+1}) \nn \\
& = &\sum_{n,m} (\frac{C_{n-m-2,2m}}{8}
 + \frac{C_{n-m,2m-2}}{2} )X^{n-m} (Y^2)^{m}
\eea
From (\ref{eq:series}) one obtains the recursive relation
\beq
\frac{C_{n-j-2,2j}}{8}  + \frac{C_{n-j,2j-2}}{2} = (-1)^{j} q_{n-1}
 {n \choose j}
\eeq
which gives
\beq
C_{n-2j,2j}= 8(-1)^{j} q_{n-j+1}{n-j+2 \choose j} - 4 C_{n-2j+2,2j-2}
\eeq
The solution of the above equation determines the wanted coefficients
\bea
C_{n-2j,2j}&=& 8(-1)^{j} q_{n-j+1}{n-j+2 \choose j} -
 8\cdot 4(-1)^{j-1} q_{n-j+2}{n-j+3 \choose j-1} + \ldots \nn\\
&=& 8(-1)^{j} \sum_{k=0}^j  4^{j-k} {n+2-k \choose k} q_{n+1-k}
\label{Ccoeff}
\eea

Now, if for the $\cN =2$ case, we define in complete analogy with what
we have done so far
\beq
\Hat{\cal X}=-({\cal K} +\bar{\cal K}) \qquad\qquad\qquad
 \Hat{\cal Y}=-\frac{i}{2}
({\cal K}-\bar{\cal K})
\eeq
we can write the non abelian generalization of the supersymmetric $\cN =2$
action in the form
\bea \label{N2nonabBI}
S_{BI}^{(2)} & = & \frac{1}{2}\int {\rm d}^4x~{\rm d}\th^4~ \Tr(\cW^2)
+ \frac{1}{2}\int{\rm d}^4x ~{\rm d}\bar{\th}^4~\Tr(\olcW^2) \\
&& + \sum_{n,m} \frac{C_{n,m}}{2} \int {\rm d}^4x~{\rm d}\th^4
~{\rm d}\bar{\th}^4 ~\STr(\cW,\cW,\olcW,\olcW,
(\Hat{\cal X})^n,(\Hat{\cal Y})^m) \nn
\eea
with the $C_{n,m}$ coefficients given in (\ref{Ccoeff}).

\vspace{0.8cm}

Although the action in (\ref{N2nonabBI}) is explicit, the computation
of the interaction terms becomes quite cumbersome as soon as one goes
to higher order in the superfield expansion.

Here we concentrate on the quartic terms and express them in terms of
$\cN =1$ superfields. Then we will quantize the action in $\cN =1$
superspace and compute perturbative corrections to the ${\cal
O}(\a'^4)$ in the spirit of \cite{DSZ}.
The quartic interaction is given by
\beq
L_4 = \frac{1}{3}  \Tr[\cW\cW\olcW\olcW
 + \frac{1}{2} \cW\olcW\cW\olcW]
\label{quartic}
\eeq
The reduction to $\cN =1$ superspace is obtained by projection  performed
now in terms of gauge covariant derivatives \cite{WGR}.
The commutator algebra is given by
\bea \label{eq:algebra}
\{\DC_{a\a},\DC_{b\b}\} = iC_{ab} C_{\a\b} \olcW && \{\DCb_{\ad}^a,
\DCb_{\bd}^b\} = iC^{ab} C_{\ad\bd} \cW \nn \\
 \{\DC_{a\a},\DCb_{\bd}^b\} = i\d^b_a \DC_{\a\bd} &&
\eea
where the  superfields $\cW$ and $\olcW$ satisfy the covariant constraints
\beq \label{eq:bianchi}
\DC^{\a}_a\DC_{b\a} \cW = C_{ac}C_{bd}\DCb^{d\ad}\DCb_{\ad}^c\olcW
\eeq
The $\cN =1$ projections are
\beq
\f \equiv \cW\vert \qq W_\a \equiv -\DC_{2\a} \cW\vert \qq
(\DC_2)^2 \cW\vert = (\DCb^1)^2 \olcW\vert = (\DCb)^2 \fb
\eeq
%%%%%%%%%%%
It is rather straightforward to obtain
\bea
&&\Tr(\DC_2)^2(\DCb_2)^2 (\cW\cW\olcW\olcW)\vert = \nn \\
&& \Tr\{W^\a W_{\a} \Wb^{\ad}\Wb_{\ad} 
        -iW^\a \f \DC_{\a\ad}\fb \Wb^{\ad}- iW^\a \f \Wb^{\ad} \DC_{\a\ad}\fb 
        -i\f W^\a \DC_{\a\ad}\fb \Wb^{\ad}\nn \\
&&      - i\f W^\a \Wb^{\ad} \DC_{\a\ad}\fb +\DCb^2\fb \f \Wb^{\ad} \Wb_{\ad} 
        +\f\DCb^2\fb \Wb^{\ad}\Wb_{\ad}+W^\a W_\a \DC^2\f\fb \nn \\
&&      + W^\a W_\a \fb\DC^2\f -i W^\a \f\fb \DC_{\a\ad} \Wb^{\ad} - i\f W^\a\fb \DC_{\a\ad} \Wb^{\ad}
        -i W^\a\f\DC_{\a\ad} \Wb^{\ad}\fb\nn \\
&&      -i\f W^\a \DC_{\a\ad} \Wb^{\ad}\fb 
        + \DCb^2\fb\f \DC^2\f\fb + \f\DCb^2\fb \DC^2\f\fb
        +\DCb^2\fb \f\fb\DC^2\f  \nn \\
&&      + \f\DCb^2\fb\fb \DC^2\f + \f\f \Box\fb \fb +\f\f \fb\Box\fb 
        +\f\f \DC^{\a\ad}\fb\DC_{\a\ad}\fb   \nn \\
%%%%%%%%%
&&   \nn \\
%%%%%%%%
&& - \frac{3}{2} i\f\f\DCb_{\ad} \Wb^{\ad} \fb\fb+
\frac{3}{2} i\f\f\fb\fb\DCb_{\ad}\Wb^{\ad}
 +i \f\f \DCb^{\ad}\fb\Wb_{\ad}\fb +i \f\f\fb \Wb_{\ad}\DCb^{\ad}\fb \nn\\
&& -i\f\f\DCb_{\ad}\fb\fb\Wb^{\ad} - i \f\f\Wb^{\ad}\fb\DCb_{\ad}\fb
   +i W^\a\DC_\a\f \f\fb\fb +i \DC_\a\f W^\a\fb\fb\f \nn \\
&& -i \f W^\a\DC_\a\f\fb \fb + i W^\a \f\fb\fb\DC_\a\f
 +\f\f[\fb,[\fb,\f]]\fb + \f\f\fb [\fb,[\fb,\f]] \}
 \label{w2}
\eea
The last set of terms is zero in the abelian case (cf.\cite{DSZ}.)
In the same way one has
\bea
&&\Tr\{\frac{1}{2}(\DC_2)^2(\DCb_2)^2 (\cW\olcW\cW\olcW)\vert\}= \nn \\
&& \Tr\{-\frac{1}{2} W^\a \Wb^{\ad} W_\a \Wb_{\ad} -iW^\a \Wb^{\ad} \f \DC_{\a\ad}\fb -i W^{\a}\DC_{\a\ad} \fb \f
   \Wb^{\ad} \nn \\
&& + W^\a \DC^2\f  W_\a\fb +\DCb^2 \fb \Wb^{\ad} \f \Wb_{\ad}
-iW^\a  \DC_{\a\ad} \Wb^{\ad} \f\fb -i W^\a\fb\f \DC_{\a\ad} \Wb^{\ad} \nn \\
&& \DCb^2\fb\DC^2\f\f\fb + \f \DC^2\f \DCb^2\fb\fb
   +\frac{1}{2}\f\DC_{\a\ad}\fb\f\DC^{\a\ad}\fb+\frac{1}{2}\f
\DC^{\a\ad}\DC_{\a\ad}\fb\f\fb  \nn\\
&& \nn \\
&& +i W^\a\fb\DC_\a\f\f\fb -i W^\a \DC_\a\f\fb \f\fb
   +i\f \fb\DC^\a\f W_\a\fb -i\f \DC^\a\f\fb W_\a\fb  \nn \\
&& +\frac{i}{2} \f\fb\f\fb\DC^\a W_\a - \frac{i}{2} \f\fb\f \DC^\a W_\a \fb
   +i\f\fb\f \DCb^{\ad}\fb\Wb_{\ad} +i\f\fb\f \Wb_{\ad}
\DCb^{\ad}\fb \nn \\
&& -i\f\fb\DCb^{\ad}\fb\f \Wb_{\ad}+i\f\DCb^{\ad}\fb\fb\f \Wb_{\ad}
   +\f (\fb\fb\f + \f\fb\fb -2\fb\f\fb) \f\fb\}
\label{W2}
\eea
In the abelian case the last set of terms vanishes (cf.\cite{DSZ}).
One can check that the quartic terms which contain only the $W^\a$
superfields reduce to the ones in (\ref{4WWbar}).

\vspace{0.8cm}

In the last part of the paper we want to perform for the non abelian
theory the same one-loop calculation performed in ref. \cite{DSZ},
i.e. we compute the ${\cal O}(\a'^4)$ on-shell divergent contributions
to the effective action with four external vector field-strengths.
The superspace $D$-algebra manipulations and the computation of the
momentum integrals for the three different types of diagrams relevant
for this calculations are exactly the same as in the abelian case.
We do not repeat here those steps. For that part of the calculation
we simply make reference to the above mentioned paper and
directly use the results obtained therein.
Thus we concentrate on
the non abelian group structure that arises in this new case.

The quartic vertices that enter the three graphs,
$G_1$, $G_2$, $G_3$ as in \cite{DSZ}, arise from the $L_4$
lagrangian in (\ref{quartic}) where we write
${\cal W}= {\cal W}^a T^a$ being $T^a$ the matrices of the gauge group.
 The colour structure associated to
a vertex  from (\ref{quartic}) is then given by
\beq
\Tr(T^a T^b T^c T^d + \frac{1}{2}T^a T^c T^b T^d)
\eeq
It is rather straightforward to examine the contributions for the
three different diagrams.
They all have a bubble type topology with two external vector lines
at each vertex. The interaction vertices are written explicitly in
(\ref{w2}) and (\ref{W2}). $G_1$ and $G_2$ have both internal quantum vector lines,
while $G_3$ contains quantum chiral fields.
One has to use the fact that the
propagators are diagonal in the colour indices and  to
take correctly into account all the various possibilities for the
contractions of the internal quantum lines.
One finds that all the three diagrams give rise to a colour factor of
the form
\beq
R(a,b;i,j) = (3!)^2~\STr(T^a T^b T^c T^d)~\STr(T^c T^d T^i T^j)
\label{R}
\eeq
where $a,b$, $i,j$ are the colour indices on the external fields at
vertex $1$ and vertex $2$ respectively.

We list the answers obtained at this stage
for the three graphs separately, using a self explanatory notation
\bea
\C^{(4)}_{G_1}[W,\Wb] &=&- \frac{1}{3^2}~
\frac{1}{2}\ub{W^{\a a}(-p_1) W_{\a}^b(-p_2)}_1~
\ub{\Wb^{\ad i}(-p_3) \Wb_{\ad}^j(-p_4)}_2 \nn \\
&& ~~{\big <}\Wb^{\bd}(1)W^{\b}(2){\big >}{\big <}
\Wb_{\bd}(1)W_{\b}(2){\big >} \cdot R(a,b;i,j)
\label{G1}
\eea
\bea
\C^{(4)}_{G_2}[W,\Wb] &=& \frac{1}{3^2}~\ub{W_{\a}^a(-p_1)
 \Wb_{\ad}^i(-p_3)}_1 ~\ub{W_{\b}^b(-p_2) \Wb_{\bd}^j(-p_4)}_2~ \nn \\
&& {\big <} W^{\a}(1)\Wb^{\bd}(2){\big >}{\big <}\Wb^{\ad}(1)
W^{\b}(2){\big >} \cdot R(a,i;b,j) \nn \\
&&+ ~~(p_3 \Leftrightarrow p_4 ; i \Leftrightarrow j)
\label{G2}
\eea
\bea
\C^{(4)}_{G_3}[W,\Wb] &=& \frac{1}{3^2}~
 \ub{W^{\a a}(-p_1)\Wb^{\ad i}(-p_3)}_1 \ub{W^{\b b}(-p_2
 )\Wb^{\bd j}(-p_4)}_2  ~\cdot \nn \\
&& \oc \f(1)i\pa_{\b\bd}\fb(2)\cc \oc i\pa_{\a\ad}\fb(1)\f(2)\cc
\cdot R(a,i;b,j) \nn \\
&&  + (p_3 \Leftrightarrow p_4 ~;~i \Leftrightarrow j)
\label{G3}
\eea
Then we perform the $D$-algebra and extract the divergent part from
the momentum integrals exactly as in \cite{DSZ} and obtain:\\
For the diagram $G_1$
\bea
\label{divG1}
\C^{(4)}_{G_1~div.} \left[ W, \olW \right] &=&
        \frac{\a^4}{18}~ \frac 1\ep~ \frac{1}{(4\pi)^2}~
        \int d^4 p_1 \dots d^4 p_4
        ~ \d \left( \S p_i \right)  d^2 \th d^2\olth~R(a,b;i,j)
                \nn\\
&& \qquad s^2 \left[ W^{\a a} (-p_1) W_\a^b (-p_2)
 \Wb^{\ad i}(-p_3) \Wb_{\ad}^j (-p_4) \right]
\eea
with $s=(p_1+p_2)^2$.\\
For the diagram $G_2$
\bea
\label{divG2}
 \C^{(4)}_{G_2~div} \left[ W, \olW \right]& =&
 \frac{\a^4}{18}~\frac{1}{2}
        ~\frac{1}{\ep} ~\frac{1}{(4\pi)^2} ~
        \int d^4 p_1 \dots d^4 p_4
        \d \left( \S p_i \right)  d^2 \th d^2\olth ~
     ~R(a,i;b,j) \nn \\
&& \qquad \quad t^2~\l[W^{\a a} (-p_1)W_\a^b (-p_2)\olW^{\ad i} (-p_3)
 \olW_{\ad}^j (-p_4)\r] \nn \\
&&\qq + (t^2 \Leftrightarrow u^2 ; i \Leftrightarrow j)
\eea
where $t = (p_1+p_3)^2$ and $u = (p_1+p_4)^2$. \\
For the diagram $G_3$
\bea
 \C^{(4)}_{G_3~div} \left[ W, \olW \right]& =&
\frac{\a^4}{18}~ \frac{1}{6}
        ~\frac{1}{\ep} ~\frac{1}{(4\pi)^2} ~
        \int d^4 p_1 \dots d^4 p_4
        \d \left( \S p_i \right)  d^2 \th d^2\olth ~R(a,i;b,j) \nn \\
&& \qq \quad t^2~ \l[W^{\a a} (-p_1)W_\a^b (-p_2)
 \olW^{\ad i} (-p_3)\olW_{\ad}^j (-p_4)\r] \nn \\
&& \qq    + (t^2 \Leftrightarrow u^2 ; i \Leftrightarrow j)
\label{divG3}
\eea
Finally the complete result is
\bea
 \C^{(4)}_{div} \left[ W, \olW \right] &= & \frac{\a^4}{18}
        ~\frac{1}{\ep} ~\frac{1}{(4\pi)^2} ~
        \int d^4 p_1 \dots d^4 p_4
        \d \left( \S p_i \right)  d^2 \th d^2\olth  \nn \\
&& \quad ~ \l[W^{\a a} (-p_1)W_\a^b (-p_2)
\olW^{\ad i} (-p_3)\olW_{\ad}^j (-p_4)\r] \nn \\
&&~~~~ \l[s^2 R(a,b;i,j) + \frac{2}{3} t^2 R(a,i;b,j)
+\frac{2}{3} u^2 R(a,j;b,i) \r]
\label{finalN2div}
\eea
The above expression can be rewritten in configuration space and made
more explicit e.g. computing the $R(a,b;i,j)$ factors for a
specific gauge group. These manipulations amount to rather simple
exercises.

Now we concentrate on the
extension of the above result to the ${\cal N}=4$ case.

\vspace{0.8cm}

As already mentioned the complete ${\cal N}=4$ supersymmetric Born Infeld
action is still unknown. What is available  are the quartic vertices of
the abelian action
\cite{T3} written in terms of ${\cal N}=1$ superfields. As compared
to the ${\cal N}=2$ case there appear three chiral superfields
instead of one. For the one-loop calculation we have reported the
only change will be in the diagram $G_3$
which contains chiral superfields propagating in the
loop. In order to obtain the corresponding ${\cal N}=4$ contribution
 it is sufficient to multiply by three the result in (\ref{divG3}),
 while the contributions from the $G_1$ and $G_2$ graphs
do not change. Thus summing (\ref{divG1}),
(\ref{divG2}) and three times (\ref{divG3})
we obtain
\bea
\label{divN4}
\C^{(4)}_{div}[W,\Wb]
&=& \frac{\a^4}{18}
        ~\frac{1}{\ep} ~\frac{1}{(4\pi)^2} ~
        \int d^4 p_1 \dots d^4 p_4
        \d \left( \S p_i \right)  d^2 \th d^2\olth  \nn \\
&& ~~~ \l[W^{\a a} (-p_1)W_\a^b (-p_2)  \olW^{\ad i} (-p_3)
\olW_{\ad}^j (-p_4)\r] \nn \\
&&~~~ \l[s^2 R(a,b;i,j) + t^2 R(a,i;b,j) + u^2 R(a,j;b,i) \r]
\eea
As compared to the ${\cal N}=2$ result we have now the factor
\begin{equation}
  \l[s^2 R(a,b;i,j) + t^2 R(a,i;b,j) + u^2 R(a,j;b,i) \r] \nn
\end{equation}
which contains more symmetries than before, e.g.
$b \leftrightarrow i ~~~~ p_2 \leftrightarrow p_3 $.\\
It is quite interesting to extract from the result in (\ref{divN4})
its bosonic component \mbox{content}. After some not so simple algebra one
obtains
\bea
\label{divN4bis}
\C^{(4)}_{div}[F]
&=& \frac{\a^4}{18}
        ~\frac{1}{\ep} ~\frac{1}{(4\pi)^2} ~
        \int d^4 p_1 \dots d^4 p_4
        \d \left( \S p_i \right)   \nn \\
&&\frac{1}{96} (t_8)^{\m_1\n_1\m_2\n_2\m_3\n_3\m_4\n_4}
F_{\m_1\n_1}^{a}(-p_1)F_{\m_2\n_2}^{b}(-p_2)
F_{\m_3\n_3}^{i}(-p_3)F_{\m_4\n_4}^{j}(-p_4) \nn \\
&& \l[s^2 R(a,b;i,j) + t^2 R(a,i;b,j) + u^2 R(a,j;b,i) \r]
\eea
where we have introduced the tensor \cite{green}
\begin{eqnarray}
&& t_8^{\m_1\n_1\m_2\n_2\m_3\n_3\m_4\n_4} =
 -\frac{1}{2}\ep^{\m_1\n_1\m_2\n_2\m_3\n_3\m_4\n_4}\nn \\
&& -\frac{1}{2}[(\d^{\m_1\m_2}\d^{\n_1\n_2}
   -\d^{\m_1\n_2}\d^{\n_1\m_2})
   (\d^{\m_3\m_4}\d^{\n_3\n_4}
   -\d^{\m_3\n_4}\d^{\n_3\m_4}) + \nn \\
&& \qq (\d^{\m_2\m_3}\d^{\n_2\n_3}
   -\d^{\m_2\n_3}\d^{\n_2\m_3})
   (\d^{\m_4\m_1}\d^{\n_4\n_1}
   -\d^{\m_4\n_1}\d^{\n_4\m_1})+\nn \\
&& \qq (\d^{\m_1\m_3}\d^{\n_1\n_3}
   -\d^{\m_1\n_3}\d^{\n_1\m_3})
   (\d^{\m_2\m_4}\d^{\n_2\n_4}
   -\d^{\m_2\n_4}\d^{\n_2\m_4})] +\nn\\
&& +\frac{1}{2}[ \d^{\n_1\m_2}\d^{\n_2\m_3}\d^{\n_3\m_4}\d^{\n_4\m_1}
   +\d^{\n_1\m_3}\d^{\n_3\m_2}\d^{\n_2\m_4}\d^{\n_4\m_1}
   + \d^{\n_1\m_3}\d^{\n_3\m_4}\d^{\n_4\m_2}\d^{\n_2\m_1}
   + \nn\\
&& + \textrm{45 antisymmetrization}]
\end{eqnarray}
In order to streamline the notation we define $F^a_{\m\n}(-p_1)\equiv
F^a_{\m\n}$ and similarly \mbox{$(b,p_2\rightarrow b)$}, $(i,p_3\rightarrow
i)$, $(j,p_4\rightarrow j)$.
Moreover for a cyclic contraction we use the notation
$F^{abij}\equiv(F^a)^{\m\n}(F^b)_{\n\rho} (F^i)^{\rho\s}(F^j)_{\s\m}$.
In this way we can write
\begin{eqnarray}
&&(t_8)^{\m_1\n_1\m_2\n_2\m_3\n_3\m_4\n_4}
  F_{\m_1\n_1}^{a}F_{\m_2\n_2}^{b}F_{\m_3\n_3}^{i}F_{\m_4\n_4}^{j}
  = 8\left[ F^{abij}+ F^{aijb}+ F^{ajbi} \right. \nn \\
&&~~~\l. -\frac{1}{4}
     (F^{a}\cdot F^{b})(F^{i}\cdot F^{j})
      -\frac{1}{4}(F^{a}\cdot F^{i})(F^{j}\cdot F^{b})-
      \frac{1}{4}(F^{a}\cdot F^{j})(F^{b}\cdot F^{i})\r]
\end{eqnarray}
which in the abelian case reduces to $4!  (F^4 - \frac{1}{4}(F^2)^2)$.\\
We can also express our result in configuration space; for example
\begin{eqnarray}
&& s^2~ R(a,b;i,j)~(t_8)^{\m_1\n_1\m_2\n_2\m_3\n_3\m_4\n_4}
  F_{\m_1\n_1}^{a}F_{\m_2\n_2}^{b}F_{\m_3\n_3}^{i}F_{\m_4\n_4}^{j}
  \Longrightarrow \nn \\
&& \Longrightarrow 32~ R(a,b;i,j)[2 (\pa^\m\pa^\n F^a)(\pa_\m\pa_\n F^b)
 F^i F^j + (\pa^\m\pa^\n F^a) F^i(\pa_\m\pa_\n F^b) F^j \nn \\
&& -\frac{1}{4}(\pa^\m\pa^\n F^a \cdot \pa_\m\pa_\n F^b)(F^i\cdot F^j)
 -\frac{1}{2}(\pa^\m\pa^\n F^a \cdot F^i)(\pa_\m\pa_\n F^b \cdot F^j)]
\end{eqnarray}
Than it becomes apparent that the three contributions in (\ref{divN4bis}) proportional to
$s^2$, $t^2$ and $u^2$ are equal, so that in configuration space the 
 final result can be written in a rather simple form
\bea
\label{nonabeldiv}
\C^{(4)}_{div}[F]
&=& \frac{\a^4}{18}
        ~\frac{1}{\ep} ~\frac{1}{(4\pi)^2} ~
        \int d^4 x
         R(a,b;i,j)  \\
&&  [2 (\pa^\m\pa^\n F^a)(\pa_\m\pa_\n F^b) F^i F^j +
 (\pa^\m\pa^\n F^a) F^i(\pa_\m\pa_\n F^b) F^j \nn \\
&& -\frac{1}{4}(\pa^\m\pa^\n F^a \cdot \pa_\m\pa_\n F^b)(F^i\cdot F^j)
-\frac{1}{2}(\pa^\m\pa^\n F^a \cdot F^i)(\pa_\m\pa_\n F^b \cdot F^j)] \nn
\eea
The above result, when restricted to abelian fields, coincides with
corresponding results in \cite{Shmakova}, \cite{DSZ} and it is
consistent with scattering amplitude calculations in type IIB string
theory on the $D3$-brane \cite{HK} and in type I open string theory
\cite{S,GW,AT}. It would be interesting to confront the non abelian
structure obtained in (\ref{nonabeldiv}) with corresponding
calculations for scattering of strings on $N$ coinciding $D$-branes
along the lines of refs. \cite{HK,GM}.

\vspace{1.3cm}
\noindent
After completion of this work the paper in ref. \cite{ketovult} has
appeared; it contains a derivation of the non abelian ${\cal N}=2$
Born Infeld action from partial breaking of ${\cal N}=4$ supersymmetry.\\

\vspace{1.1cm}

Acknowledgements\\
This work has been partially supported by INFN, MURST and the
European Commission TMR program
ERBFMRX-CT96-0045, in which N. T. and D. Z. are associated to the
University of Torino.

\newpage
\begin{flushleft}

\end{flushleft}
\end{document}